\documentclass[traditabstract]{aa}
\usepackage{graphicx}
\usepackage{rotating,subfigure,amssymb,afterpage}
\usepackage{txfonts}
\usepackage{natbib}
\usepackage[pdftitle={Local Underdensity in the{\sf REFLEX II} Survey},
pdfauthor={H. B\"ohringer et al.},
pdfsubject={}, pdfkeywords={}, bookmarks, bookmarksopen, 
colorlinks=true, linkcolor=blue, citecolor=blue, anchorcolor=blue,
urlcolor=blue]{hyperref} 
\newcommand{\propsim}{\lower 3pt \hbox{$\, \buildrel {\textstyle
      \propto}\over {\textstyle \sim}\,$}}
\begin{document}
   \title{The extended ROSAT-ESO Flux-Limited X-ray Galaxy Cluster
   Survey (REFLEX II)\\ V. Exploring a local underdensity in the
   Southern Sky
\thanks{
   Based on observations at the European Southern Observatory La Silla,
   Chile}}

   \author{Hans B\"ohringer\inst{1}, Gayoung Chon\inst{1}, 
       Martyn Bristow\inst{2}, Chris A. Collins\inst{2}}

   \offprints{H. B\"ohringer, hxb@mpe.mpg.de}

   \institute{$^1$ Max-Planck-Institut f\"ur extraterrestrische Physik,
                   D-85748 Garching, Germany.\\
              $^2$ Astrophysics Research Institute, Liverpool John Moores University, 
                    IC2, Liverpool Science Park, 146 Brownlow Hill,
                    Liverpool L3 5RF, UK
}

   \date{Submitted 15/8/14}

\abstract{Several claims have been made that we are located 
in a locally underdense region of the Universe based on observations 
of supernovae and galaxy density distributions. Two recent 
studies of K-band galaxy surveys have, in particular, provided new 
support for a local underdensity in the galaxy distribution out 
to distances of 200 - 300 Mpc. If confirmed, such large local 
underdensities would have important implications on the interpretation
of local measurements of cosmological parameters. Galaxy clusters 
have been shown to be ideal probes to trace the large-scale structure 
of the Universe. In this paper we study the local density distribution 
in the southern sky with the X-ray detected galaxy clusters from 
the {\sf REFLEX II} cluster survey. From the normalized comoving number 
density of clusters we find an average underdensity 
of $\sim 30 - 40\%$ in the redshift range out to $z \sim 0.04$ 
($\sim 170$ Mpc) in the southern extragalactic sky
with a significance larger than 3.4$\sigma$. On larger scales from 
300 Mpc to over 1 Gpc the density distribution appears remarkably
homogeneous. The local underdensity 
seems to be dominated by the South Galactic Cap region.
A comparison of the cluster distribution with that of 
galaxies in the K-band from a recent study 
shows that galaxies and clusters trace each other very closely
in density. In the South Galactic Cap region both surveys find a 
local underdensity in the redshift range $z= 0$ to $0.05$ and no 
significant underdensity in the North Galactic Cap at southern latitudes.
Cosmological models that attempt to interpret the cosmic acceleration 
deduced from supernova type Ia observations by a large local void without 
the need for reacceleration, require that we are located close to 
the center of a roughly spherical void with a minimal size
of $\sim 300$ Mpc. In contrast our results show that the local underdensity
is not isotropic and limited to a size significantly smaller 
than 300 Mpc radius.}

 \keywords{galaxies: clusters, cosmology: observations, 
   cosmology: large-scale structure of the Universe, 
   X-rays: galaxies: clusters} 

\authorrunning{B\"ohringer et al.}
\titlerunning{{\sf REFLEX II} Cluster X-ray Luminosity Function}
   \maketitle
%
%________________________________________________________________

\section{Introduction}

Measurements of global cosmological parameters are generally evaluated in the 
context of a homogeneous, isotropic cosmological model. This assumption is well 
supported on large scales by observations of the cosmic microwave background 
by the WMAP (Hinshaw et al. 2013) and PLANCK (PLANCK Collaboration XVI and XXIII, 2013) 
satellites, leaving little room for deviations from this model, with some 
unexplained anomalies at the less than $3\sigma$ level. But even in the 
frame of a homogeneous Universe on very large scales, another question 
remains for observational constraints of cosmological parameters: is the 
point from which we observe our Universe representative or peculiar? 
This is particularly important for measurements of cosmological parameters 
carried out in the local Universe, for example, the Hubble constant. 
This problem was recognized early e.g. by Turner et al. (1992) 
pointing out that a locally underdense Universe would yield a Hubble 
constant larger than the cosmic mean.

With observational evidence from supernovae (SN) for a accelerating universe 
(Perlmutter et al. 1999, Schmidt et al. 1998), models with local voids were 
considered as an alternative explanation to the supernovae data 
without Dark Energy or a cosmological
constant (e.g. C\'el\'erier 2000, Tomita, 2000, 2001, 
Alexander et al. 2009, February et al. 2010
and references therein). Alexander et al. (2009), for example, obtain
a minimum size for such a void of about $200 h^{-1}$ Mpc with a mean
mass density deficiency of $\sim 40\%$ to explain the SN and CMB data 
in an Einstein-de Sitter universe.
Moss et al. (2011) critically discuss the viability of these 
void models, pointing out the difficulties in reproducing 
all observational data including Baryonic Acoustic Oscillations and 
cluster abundances. More recently Marra et al. (2013) explored how 
much the tension between the 
Hubble parameter determined by Riess et al. (2011) with local SN observations of 
73.8 ($\pm 2.4$) km s$^{-1}$ Mpc$^{-1}$ and the value of 
67.3 ($\pm 1.2$) km s$^{-1}$ Mpc$^{-1}$ from PLANCK  
(Planck Collaboration XVI 2013) can be reconciled by a local void model,
rejecting a void model as not very likely.

Zehavi et al. (1998) and more recently Jha et al. (2007) found 
an indication for a local underdensity in the data of supernovae type Ia 
inside a radius of about 300 $h^{-1}$ Mpc yielding a $\sim 6.5\%$ higher 
value for the Hubble constant inside a radius of about 70 $h^{-1}$ Mpc 
compared to the region outside. In contrast, Hudson et al. (2004) 
find a local flow excess, $\Delta H / H$, of  only 2.3 ($\pm 1.9$)\%
with 98 SN and Conley et al. (2007) show for the same data
as used in Jha et al. (2007)  that the presence
of the local underdensity depends on how the SN colors are modelled.
With a sample of 76 galaxy clusters with Tully-Fisher distances for galaxies,
Giovanelli et al. (1999) characterise the local Hubble flow out to 
200 $h^{-1}$ Mpc and find a surprisingly smooth Hubble flow in the distance
range 50 - 200 $h^{-1}$ Mpc with variations smaller than 1 $\pm 2$\%.

Evidence for local voids with sizes of $\sim 200 h^{-1}$ Mpc has also 
been claimed from studies of galaxy counts and galaxy redshift surveys,
for example by Huang et al. (1997), Frith et al. (2003, 2006), 
and Busswell et al. (2004). Two more recent survey results by Keenan
et al. (2013) and Whitbourn \& Shanks (2014) have renewed the interest
in these studies. The first study examines the K-band galaxy 
luminosity function from the UKIDSS Large Area and 2MASS
Surveys with spectroscopy from SDSS, 2dFGRS, GAMA
(Galaxy And Mass Assembly, Driver et al. 2011), and 6dFGRS 
finding a density deficit of $\sim 30 -50\%$ inside a radius 
of about $300 h^{-1}$ Mpc (at redshifts $\le 0.07$).
Whitbourn \& Shanks (2014) study the galaxy density distribution 
in three larger regions in the South Galactic Cap (SGC), 
the southern part of the North Galactic Cap (NGC), and
the northern part of the NGC using 2MASS K-band magnitudes in
connection with 6dFRGS, GAMA, 
and SDSS spectroscopic data including $\sim 250 000$ galaxies 
out to $z = 0.1$. They 
find a large underdense region with a deficit of about 40\% 
inside a radius of $150 h^{-1}$ Mpc in the SGC, no
deficit in the southern part of the NGC, and a less 
pronounced underdensity in the NGC north of the
equator. These types of underdensities indicated in the
galaxy distributions - if extrapolated from the survey regions
to the entire celestial sphere -
are of about the size of the minimal void models
mentioned above. This makes these findings particularly
interesting.

Another excellent set of probes for the large-scale structure are
galaxy clusters, which constitute statistically well defined density peaks
within the large-scale matter distribution (e.g. Bardeen et al. 1986).
In this paper we are using the statistically complete sample of galaxy clusters
from the {\sf REFLEX II} galaxy cluster survey in the southern sky. The survey is 
characterised by a well defined selection function (B\"ohringer et al. 2013)
and we have used it successfully for a statistical assessment
of the large-scale structure.
In the {\sf REFLEX I} survey we already detected an indication for a lower cluster 
density in the southern sky in the redshift range $z = 0.02 - 0.04$ which we
attributed to large-scale structure (Schuecker et al. 2001). We also
observed an X-ray luminosity function in {\sf REFLEX I} with
a larger amplitude at the low luminosity end 
for the region in the southern NGC compared to the SGC
(B\"ohringer et al. 2002). The difference was consistent with the expected
cosmic variance of the survey regions and we took this result as a hint
for a lower local density in the south cap compared to the north cap
region. With the extension of the survey, {\sf REFLEX II}, comprising about twice
as many clusters than {\sf REFLEX I}, we have the best galaxy cluster sample at 
hand to probe the density distribution in the local Universe.
 
The paper is organised as follows. In section 2 we outline the properties
of the {\sf REFLEX II} cluster sample and in section 3 we describe the 
analysis methods. Section 4 provides the results on the cluster 
density distribution.
We compare our results to galaxy surveys in section 5. Section 6
provides a discussion and we close the 
paper with a summary and conclusions in section 7.
For the determination of all parameters that depend on distance we use
a flat $\Lambda$CDM cosmology with the parameters $H_0 = 70$ km s$^{-1}$
Mpc$^{-1}$ and $\Omega_m = 0.3$. Exceptions are the literature values 
quoted above with a scaling by $h = H_0/ 100$ km s$^{-1}$
Mpc$^{-1}$.

\section{The REFLEX II Galaxy Cluster Survey}

The {\sf REFLEX II} galaxy cluster survey is based on the X-ray detection of
galaxy clusters in the RASS (Tr\"umper 1993, 
Voges et al. 1999). The region of the survey is
the southern sky below equatorial latitude +2.5$^o$ and at galactic
latitude $|b_{II}| \ge 20^o$. The regions of the Magellanic clouds have
been excised. The survey region selection, the source detection, the
galaxy cluster sample definition and compilation, and the construction of
the survey selection function  as well as tests of the completeness of the
survey are described in B\"ohringer et al. (2013). In summary the 
survey area is $ \sim 4.24$ ster. The nominal flux-limit down to which
galaxy clusters have been identified in the RASS in this region is
$1.8 \times 10^{-12}$ erg s$^{-1}$ cm$^{-2}$ in the
0.1 - 2.4 keV energy band. For the assessment of the large-scale structure
in this paper we apply an additional cut
on the minimum number of detected source photons of 20 counts. This has
the effect that the nominal flux cut quoted above is only reached in about
80\% of the survey. In regions with lower exposure and higher interstellar
absorption the flux limit is accordingly higher 
(see Fig.\ 11 in B\"ohringer et al. 2013). This effect is modelled and
taken into account in the survey selection function.

We have already demonstrated with the {\sf REFLEX I} survey 
(B\"ohringer et al. 2004) that clusters provide a very precise means to 
obtain a census of the cosmic large-scale matter distribution
through e.g. the correlation function (Collins et al. 2000), 
the power spectrum (Schuecker et al. 2001, 2002, 2003a, 2003b), 
Minkowski functionals, (Kerscher et al. 2001),
and, using {\sf REFLEX II}, with the study
of superclusters (Chon et al. 2013, 2014) and the cluster power 
spectrum (Balaguera-Antolinez et al. 2011). 
The fact that clusters follow the large-scale
matter distribution in a biased way, that is with an amplified amplitude of
the density fluctuations (see e.g. Balaguera-Antolinez et al. 2011, 2012,
for measurements of the bias in the {\sf REFLEX} survey), 
is a valuable advantage, which makes it easier to detect local density
variations.

The flux limit imposed on the survey is for a nominal flux, that has been
calculated from the detected photon count rate for a cluster X-ray spectrum
characterized by a temperature of 5 keV, a metallicity of 0.3 solar,
a redshift of zero, and an interstellar absorption column
density given by the 21cm sky survey described
by Dickey and Lockmann (1990). This count rate to flux conversion is an
appropriate prior to any redshift information and is analogous to an observed
object magnitude corrected for galactic extinction in the optical.

After the redshifts have been measured, a new flux is calculated taking the
redshifted spectrum and an estimate for the spectral temperature
into account. The temperature is estimated by means of the X-ray luminosity -
temperature relation from Pratt et al. (2009) determined from the {\sf REXCESS} cluster
sample, which is a sample of clusters drawn from {\sf REFLEX I} for deeper follow-up
observations with XMM-Newton, which is representative of the entire flux-limited
survey (B\"ohringer et al. 2007). The luminosity is determined first from the
observed flux by means of the luminosity distance for a given redshift. Using
the X-ray luminosity mass relation given in Pratt et al. (2009) we can then
use the mass estimate to determine a fiducial radius of the cluster, which is
taken to be $r_{500}$ \footnote{$r_{500}$ is the radius where the average
mass density inside reaches a value of 500 times the critical density
of the Universe at the epoch of observation.}. We then use a beta model for the
cluster surface brightness distribution to correct for the possibly missing
flux in the region between the detection aperture of the source photons and
the radius $r_{500}$. The procedure to determine the flux, the luminosity,
the temperature estimate, and $r_{500}$ is done iteratively and described in
detail in B\"ohringer et al. (2013). In that paper we deduced a mean flux 
uncertainty for the {\sf REFLEX II} clusters of 20.6\%, which is 
mostly due to the Poisson statistics of the source counts but also 
contains some systematic errors. 

The X-ray source detection and selection is based on the official RASS source
catalogue by Voges et al. (1999). We have been using the publicly available
final source catalog \footnote{the RASS source catalogs can
be found at: http://www.xray.mpe.mpg.de/rosat/survey/rass-bsc/  for the
bright sources and http://www.xray.mpe.mpg.de/rosat/survey/rass-fsc/ for
the faint sources} as well as a preliminary source list that
was created while producing the public catalogue.
To improve the quality of the source parameters for the mostly
extended cluster sources, we have reanalyzed all the X-ray sources with the
growth curve analysis method (B\"ohringer et al. 2000). The flux cut was imposed
on the reanalysed data set. 

\begin{figure}
   \includegraphics[width=\columnwidth]{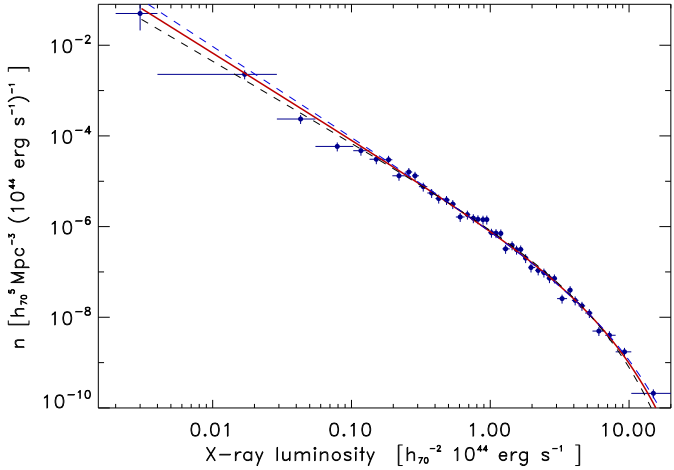}
\caption{{\sf REFLEX II} X-ray luminosity function for the redshift range
z = 0 - 0.4. We also show the best fitting Schechter function
 and two bracketing solutions indicating the estimated uncertainties.
}\label{fig1}
\end{figure}

The galaxy clusters among the sources have been identified using all 
available literature, data base information,
and finally follow-up observations at ESO La Silla. The source identification
scheme is described in detail in B\"ohringer et al. (2013). The redshifts have
been secured mostly by multi-object spectroscopy and the redshift accuracy
of the clusters is typically 60 km s$^{-1}$ (Guzzo et al. 2009, 
Chon \& B\"ohringer 2012).

The survey selection function has been determined as a function
of the sky position with an angular resolution of one degree
and as a function of redshift. The survey selection function
takes all the systematics of the RASS exposure distribution, galactic
absorption, the fiducial flux, the detection count limit, and all the applied 
corrections described above into account. The survey selection
function is a very important pre-requisite for the precise large-scale
structure assessment performed in this paper.

\section{Method}

%_________________________________One column table----------------------------
   \begin{table}
      \caption{Best fitting parameters for a Schechter function describing
       the {\sf REFLEX II} X-ray luminosity function. For the description 
       of the parameters of the Schechter function see Eq. 1.
        $L_X^{\ast}$ has units of $10^{44}$ erg s$^{-1}$ in the 0.1 - 2.4
        keV band and $n_0$ of $h_{70}^5$ Mpc$^{-3}$ ($10^{44}$ erg s$^{-1}$)$^{-1}$.}
         \label{Tempx}
      \[
         \begin{array}{llll}
            \hline
            \noalign{\smallskip}
  ~~L_x-{\rm range}& ~~~~~~\alpha  & ~~~~~~L_X^{\ast} & ~~~~~~n_0 \\
            \noalign{\smallskip}
            \hline
            \noalign{\smallskip}
{\rm best}    & 1.92  & 3.95 & 2.83\cdot 10^{-7}  \\
{\rm low}    & 1.8  & 3.2 & 4.4\cdot 10^{-7}  \\
{\rm high}    & 2.0  & 4.7 & 2.0\cdot 10^{-7} \\
            \noalign{\smallskip}
            \hline
            \noalign{\smallskip}
         \end{array}
      \]
%\begin{list}{}{}
%\item[$^{\rm a}$] ......
%\end{list}
%{\bf Notes:} 
\label{tab1}
   \end{table}
%_____________________Table-END______________________________________

Since we are dealing with a flux-limited X-ray galaxy cluster
sample with additional modulation of the survey selection
function across the sky,
we cannot directly determine the density distribution of
galaxy clusters without taking the selection function into
account. We include this correction in our study of the relative
density distribution with respect to the mean density in the following
way. We determine the predicted distribution of the galaxy
clusters in our survey, e.g. as a function of redshift, based
on the X-ray luminosity function determined in B\"ohringer et al. 
(2014) and the selection function of the {\sf REFLEX II} survey. 
The relative density variations are then determined 
by the ratio of the observed and expected number of galaxy clusters.

For an analytical description of the {\sf REFLEX} X-ray luminosity 
function we use a Schechter function of the form

\begin{equation}
{n(L_X)~dL_X}~ =~ n_0~ \left( {L_X \over L_X^{\ast}}\right)^{-\alpha}
exp\left(- {L_X \over L_X^{\ast}}\right)  {dL_X \over L_X^{\ast}} ~~~.
\end{equation}

The parameters used for the Schechter function are given in Table 1.
In addition to the best fitting function we also use two bracketing 
functions, also given in Table 1, which capture the uncertainty 
in the fit of the Schechter function parameters. The observed luminosity 
function and the three Schechter functions are also shown in Fig. 1.
In our study in B\"ohringer et al. (2014) we found no significant evolution
of the X-ray luminosity function of the {\sf REFLEX II} clusters
in the redshift interval $z = 0$ to $0.4$. Therefore we assume this
function to be constant in the volume studied here.

This approach of determining the relative density distribution of 
the clusters through the ratio of the predicted and observed systems
confronts us with an intrinsic problem. In the flux-limited 
sample, the low luminosity part of the X-ray luminosity function is
determined from the cluster population at low redshifts
where these clusters can be detected. If we have a large 
enough locally underdense region in the survey, this will 
bias the measured X-ray luminosity function low at the low 
luminosity end. Using such a biased luminosity function we cannot
properly assess the underdense region. One way to check
for this effect is the use of volume-limited subsamples.
We show in Fig. 2 the median lower limit of the X-ray luminosity
for cluster detection in {\sf REFLEX II} with 20 photons
as a function of redshift. The maximum redshift out to which a
volume-limited sample with a certain minimal X-ray luminosity
can be constructed can be read off from this plot.
For a lower limit of $L_x = 10^{42}$ erg s$^{-1}$ 
for example the maximum 
redshift is $z = 0.0156$, for $L_x = 2 \times 10^{43}$ erg~s$^{-1}$
it is $z = 0.068$. We use this information below to check the 
results of the purely flux-limited approach. We also show 
the boundaries of six volume-limited subsamples of
{\sf REFLEX II} that are also used below to assess the 
cluster density distribution.

\begin{figure}
   \includegraphics[width=\columnwidth]{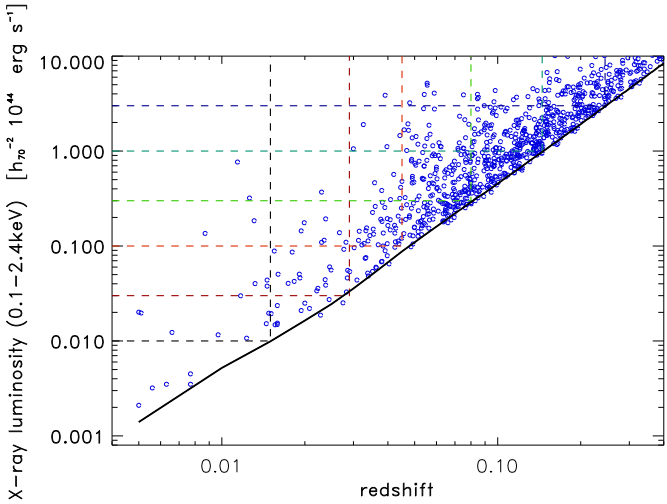}
\caption{Median lower detection limit for the X-ray luminosity as a function
of redshift in the flux-limited {\sf REFLEX II} survey (black line).
The data points show the luminosity and redshift distribution of the 
individual {\sf REFLEX II} clusters.
We also show the limits of six volume-limited subsamples of the  {\sf REFLEX II}
survey used to assess the density distribution in this paper.
}\label{fig2}
\end{figure}

\section{Results}

\begin{figure}
   \includegraphics[width=\columnwidth]{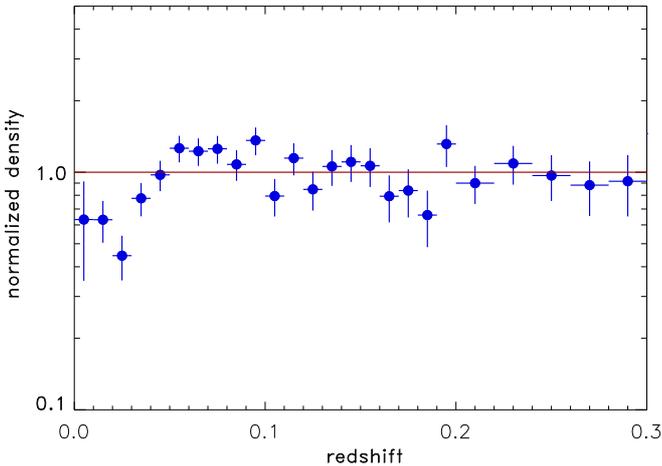}
\caption{Cluster density distribution as a function of 
redshift for the {\sf REFLEX II} galaxy clusters for a minimum luminosity of 
$10^{42}$ erg s$^{-1}$ (0.1 - 2.4 keV). The density distribution has been
normalised by the predicted redshift distribution as explained in the text.
}\label{fig3}
\end{figure}

In Fig. 3 we show the density ratio distribution of the clusters
for the entire {\sf REFLEX II} cluster sample with 
$L_x \ge 10^{42}$ erg s$^{-1}$ out to a redshift of $z = 0.3$.
It was constructed by dividing
the observed number of {\sf REFLEX II} clusters in different redshift bins 
by the prediction based on the best fitting Schechter and 
{\sf REFLEX II} selection function. 203 clusters are involved in tracing the
density at $z \le 0.06$ and 416 in the region out to $z = 0.1$.
While the overall cluster distribution 
is remarkably homogeneous, we note an underdensity of 
about 30 - 40\% at $z \le 0.04$ followed by an overdensity 
from $z = 0.04$ to $z = 0.07$.

To illustrate the point made above, that the determined
density ratio distribution depends, to some extent, on the luminosity
function adopted, we show in Fig. 4 the density ratio distributions
for the three different luminosity functions listed in Table 1.
For the X-ray luminosity function that is biased low at the low 
luminosity end, the underdensity effect weakens, as expected, 
but the overdensity in the redshift range $z = 0.04 - 0.07$
becomes slightly more pronounced. However, there is still a significant density
difference between the low density and higher density regions
at low redshifts. For the function with a large bias at the 
low luminosity end the depth of the local underdensity increases. 

\begin{figure}
   \includegraphics[width=\columnwidth]{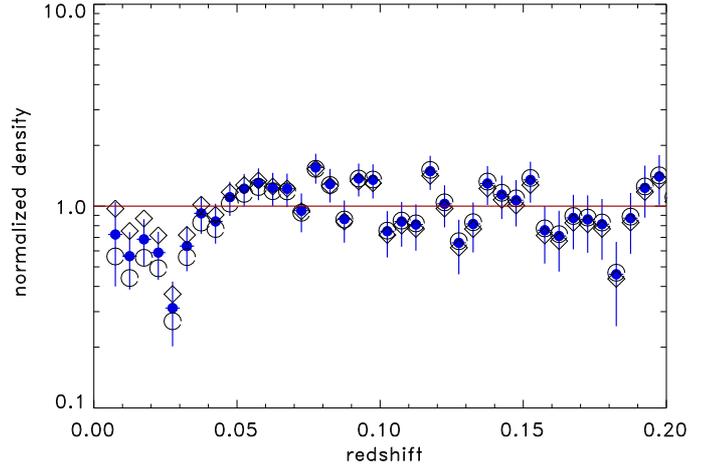}
\caption{Cluster density distribution as a function of 
redshift for the {\sf REFLEX II} galaxy clusters for a minimum luminosity of 
$10^{42}$ erg s$^{-1}$ (0.1 - 2.4 keV). Same as Fig. 3, adding the results for
using the bracketing X-ray luminosity functions: low bias (open diamonds)
and high bias (open circles).
}\label{fig4}
\end{figure}

To further test the robustness of the underdensity, 
we use a larger value of the
lower X-ray luminosity limit of the sample. We show in
Fig. 5 the density ratio distribution for a lower
luminosity limit of $0.2  \times 10^{44}$ erg s$^{-1}$
for which the cluster sample is volume-limited up
to a redshift of $z = 0.068$. We clearly detect the 
previously identified underdense and overdense regions
within the range where the sample is volume-limited.

\begin{figure}
   \includegraphics[width=\columnwidth]{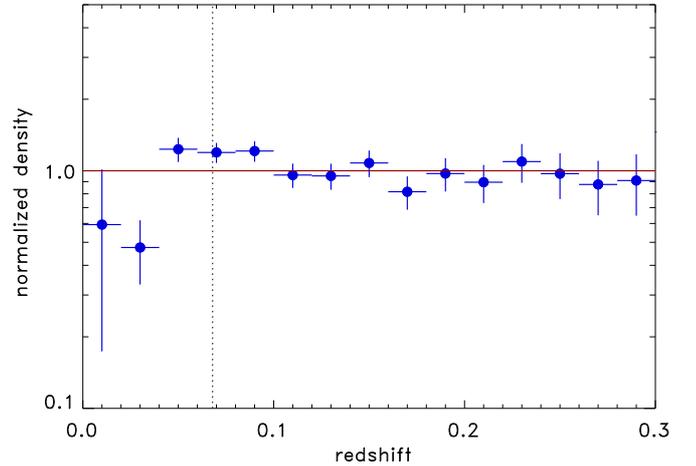}
\caption{Cluster density distribution as a function of 
redshift for the {\sf REFLEX II} galaxy clusters for a minimum luminosity of 
$0.2 \times 10^{44}$ erg s$^{-1}$ (0.1 - 2.4 keV). The density distribution has been
normalised by the predicted redshift distribution as explained in the text.
The vertical dotted line shows the redshift range up to which the sample
is essentially volume-limited.
}\label{fig5}
\end{figure}

%_________________________________One column table----------------------------
   \begin{table}
      \caption{Comparison of the ratio of the observed to 
predicted cumulative number of clusters as a function of 
the redshift limit for the {\sf REFLEX II} clusters sample.
We also give the statistical uncertainty in brackets.
The columns labeled (1) to (3) give
the ratio of observed to predicted numbers for the 
{\sf REFLEX II} sample for a lower X-ray luminosity limit of 
$L_X = 10^{42}$ erg s$^{-1}$,
for the best fitting, low, and high X-ray luminosity function.
The last column, (4), shows the results for the {\sf REFLEX II} 
sample with an X-ray luminosity limit of 
$L_X = 0.2 \times 10^{44}$ erg s$^{-1}$.}
         \label{Tempx}
      \[
         \begin{array}{lllll}
            \hline
            \noalign{\smallskip}
  {\rm z~limit} & {\rm (1)}  & {\rm (2)} & {\rm (3)} 
         & {\rm (4) }   \\
            \noalign{\smallskip}
            \hline
            \noalign{\smallskip}
0.02  &  0.63 (\pm 0.11) & 0.82 (\pm 0.15) & 0.50 (\pm 0.09) & 0.59 (\pm 0.42)  \\
0.03  &  0.53 (\pm 0.07) & 0.67 (\pm 0.09) & 0.44 (\pm 0.06) & 0.27 (\pm 0.15)  \\
0.04  &  0.61 (\pm 0.07) & 0.73 (\pm 0.08) & 0.52 (\pm 0.05) & 0.49 (\pm 0.14)  \\
0.05  &  0.71 (\pm 0.06) & 0.83 (\pm 0.07) & 0.61 (\pm 0.05) & 0.76 (\pm 0.12)  \\
0.06  &  0.81 (\pm 0.05) & 0.93 (\pm 0.07) & 0.72 (\pm 0.05) & 1.01 (\pm 0.10)  \\
0.07  &  0.88 (\pm 0.06) & 0.98 (\pm 0.06) & 0.78 (\pm 0.05) & 1.05 (\pm 0.09)  \\
0.08  &  0.93 (\pm 0.05) & 1.02 (\pm 0.06) & 0.84 (\pm 0.05) & 1.10 (\pm 0.08)  \\
            \noalign{\smallskip}
            \hline
            \noalign{\smallskip}
         \end{array}
      \]
%\begin{list}{}{}
%\item[$^{\rm a}$] ......
%\end{list}
%{\bf Notes:} 
\label{tab1}
   \end{table}
%
%________________________Table-END______________________________________

Taking this approach one step further we show in Fig. 6 the density 
distribution traced by six different volume-limited subsamples 
obtained from {\sf REFLEX II}. The luminosity and redshift boundaries 
for the samples are shown in Fig. 2. The colors of the subsample limits
(in the electronic version of the paper) are the same as the colors
of the data points in Fig. 6.
The relative normalisations of the samples are based on weights
determined from an integration of the observed interval of the
X-ray luminosity function. These weights are given by

\begin{equation}
W_i = {\int_{L_{X_0}}^{\infty} \phi(L) dL \over \int_{L_{X_i}}^{\infty} \phi(L) dL} ~~~, 
\end{equation}

where $L_{X_0}$ is a reference lower limit and $L_{X_i}$ is the lower
X-ray luminosity limit of the sample. The reasonable agreement of
the sample densities in the overlapping redshift regions  
provides a direct test of the quality of this inter-calibration.
The different samples clearly trace the local underdensity 
and the following overdense region. Given the various tests presented,
we can be confident that these density variations are real and
not an artifact of the adopted luminosity function.

\begin{figure}
   \includegraphics[width=\columnwidth]{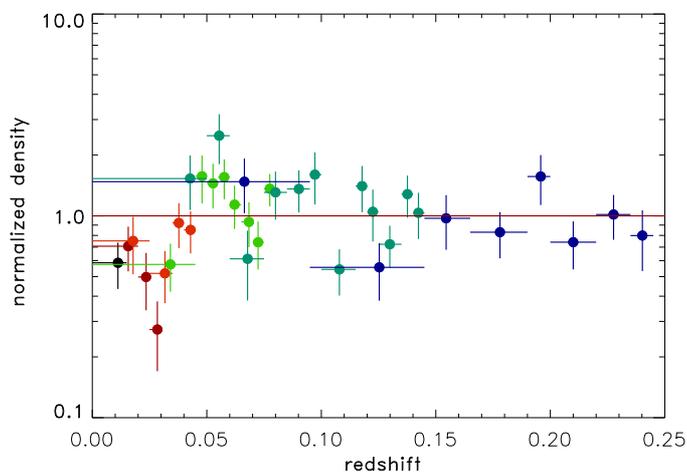}
\caption{Relative density distribution of the {\sf REFLEX II} clusters
represented by six volume-limited subsamples as a function of redshift. 
The relative normalisation of the subsamples has been determined 
from Eq. 1 as explained in the text.
}\label{fig6}
\end{figure}

From an inspection of Table 2 we conclude on the following results.
In the redshift range $z = 0$ to $0.03$ we find an underdensity in the 
southern cluster distribution of about 40 - 50\%. Taking the systematics 
of the luminosity function uncertainties into account, 
the statistical errors leave us with a detection 
confidence limit of $3.7 \sigma$ for the $L_X \ge 10^{42}$ erg s$^{-1}$
sample and a significance larger than $4 \sigma$ for the volume-limited
sample shown in Fig. 5. 
For the redshift range $z = 0 - 0.04$ we obtain an underdensity 
of about 30 - 40\% with a significance of $3.4 \sigma$ for the flux-limited
sample and $3.6 \sigma$ for the volume-limited sample.

These results are consistent with our previous findings from {\sf REFLEX I}
for an underdensity in the cluster distribution at $z < 0.04$
and an overdensity in the region $z = 0.05$ to $0.06$ (Schuecker
et al. 2001). At a closer look, the {\sf REFLEX I} results show
no underdensity at $z < 0.02$ and the overall deficit is slightly 
less pronounced. This is partly due to the fact that the low luminosity
end of the {\sf REFLEX I} X-ray luminosity function is less well sampled
yielding a slightly smaller negative slope than for {\sf REFLEX II}. 
With the new data, the 
X-ray luminosity function is better established leading to
a better assessment of the low redshift density distribution.  

To explore where the underdensity is located in the southern 
sky we study the sky distribution of all clusters at redshift
$z \le 0.06$. To correct for the different depth to which the
luminosity function is probed at different redshifts and sky
positions, we weight each cluster by the values determined
by Eq. 2, where $L_{X_i}$ is the X-ray luminosity of the detection
limit at the sky and redshift location of the clusters 
to be weighted. To limit the Poisson noise of the 
cluster distribution to a value significantly
smaller than the actual density variations we smooth the sky distribution 
with a Gaussian kernel with a $\sigma$ of 10 degrees on the sky. 

The resulting large-scale density distribution is shown in Fig. 7.
We observe three prominent regions with overdensities by about
a factor of two. The region around RA $ = 200^o$
and DEC $ = -30^o$ is dominated by the Shapley supercluster.
The region around RA $ = 300^o$ and DEC $ = -55^o$ is marked
by the supercluster 120 and the region  RA $ = 50 - 100^o$
and  DEC $ = -55^o$ is marked by superclusters 42 and 62 identified
in the {\sf REFLEX II} sample by Chon et al. (2013).
These superclusters are all located between $z = 0.04$ and 0.065
and thus are mainly responsible for the overdensity in the total
{\sf REFLEX II} sample in this redshift range as found above.
In particular the regions in the SGC at low declination 
and near the South Galactic Pole are mostly underdense.     

\begin{figure}
   \includegraphics[width=\columnwidth]{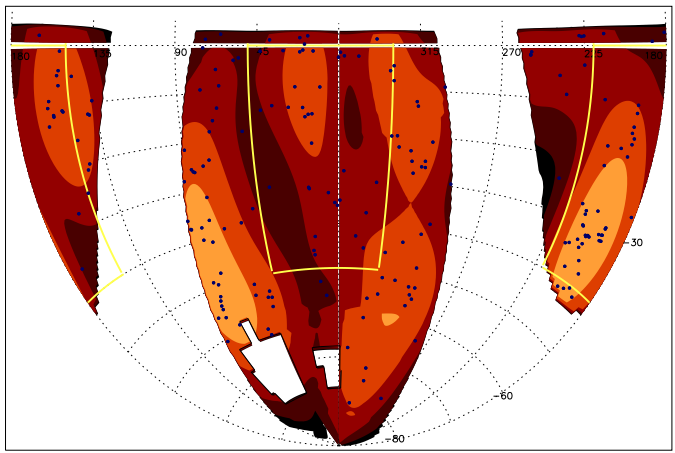}
\caption{Cluster density distribution of {\sf REFLEX II} galaxy clusters in the
redshift shell z = 0.0 - 0.06 shown in equatorial coordinates.
The density distribution was smoothed
by a Gaussian kernel with $\sigma = 10$ deg., normalised to an integral 
of 1. The color scale for the density ratio compared to the mean 
is defined by orange $ \ge 2$, light red $= 1 - 2$,
brown $= 0.5 - 1$, dark brown $ \le 0.5$. The regions of the Magellanic 
Clouds are excised from the {\sf REFLEX II} Survey and the galaxy survey regions
of Whitbourn \& Shanks (2014) are indicated by yellow solid lines.
}\label{fig7}
\end{figure}

\section{Comparison to galaxy redshift surveys}

We can compare our results in more detail with the findings from
galaxy redshift surveys of a local underdensity. We use the
galaxy cluster distribution to specifically investigate two of the 
sky regions studied by Whitbourn \& Shanks (2014) in the southern
sky by means of the 2MASS photometric and
2MASS and 6dF redhift surveys. The regions are located
in the SGC (6dF SGC, 3511 deg$^2$) at RA $= 0 - 50^o$ and 
$330 - 360^o$ ) with DEC $= -40 - 0^o$ as well as at RA $= 150 - 220^o$   
with DEC $= -50 - 0^o$  in the NGC (6dF NGC, 2578  deg$^2$) . We extract
cluster data from exactly the same regions as indicated in Fig. 7.
In Figs. 8 and 9 we show the relative redshift distribution of
the {\sf REFLEX II} clusters in  these areas. The distribution 
functions shown are again the ratio of the observed to 
predicted galaxy clusters for the best fitting X-ray luminosity 
function. We note the underdensity in the redshift region 
$z = 0$ to $0.05$ (corresponding to $\sim 212$ Mpc) in the SGC region,
while the NGC shows no such 
significant underdensity. The plots also show the density
distributions determined by Whitbourn \& Shanks (2014) for
the 6dF-SGC and 6df-NGC regions as a function of redshift
out to $z = 0.1$. The correspondence of the galaxy and cluster
distributions are suprisingly close. Peaks and troughs are traced 
by both distribtions in almost the same way.

A comparison of our galaxy cluster distribution with the 
results of Keenan et al. (2013) is more difficult, since their
survey only covers a total area of $\sim 585.4$ deg$^2$
with three different subregions of which two overlap with
the {\sf REFLEX} survey. These two regions cover a band with a width
of $\pm 2$ degrees around the equator. The region in the 
SGC stretches roughly from RA of 300 to 80 degrees and
the one in the NGC from about RA of 130 to 260 degrees. 
Since this sky area is much smaller than that of the former 
study, the cluster number statistics is too poor to precisely
trace the density distribution. To improve the statistics
at least by some factors, we look at the cluster distribution 
in a latitude band in the declination range  -4$^o$ to +2.5$^o$. 
In the entire region in this declination
range covered by {\sf REFLEX} we cannot detect any significant depression
in the density distribution at low redshift. If we, however,
limit our study to the right ascension region of the SGC 
given above, we find an underdensity in the redshift
range $z = 0.0$ to $0.075$ and a peak at  $z = 0.08$ to $0.1$.
This is similar to the observations of Keenan et al. for the SGC.
The density peak at $z \sim 0.09$ is connected to the region 
containing the Sloan Great Wall. For the NGC region
we do not observe a low density at low redshift, but an indication
for underdense regions at $z = 0.04$ to $0.05$ and  
at $z = 0.06$ to $0.08$. Thus there is some similarity of the
results. But the comparison is based on small number statistics
for the clusters and we cannot expect a precise correspondence. 

\begin{figure}
   \includegraphics[width=\columnwidth]{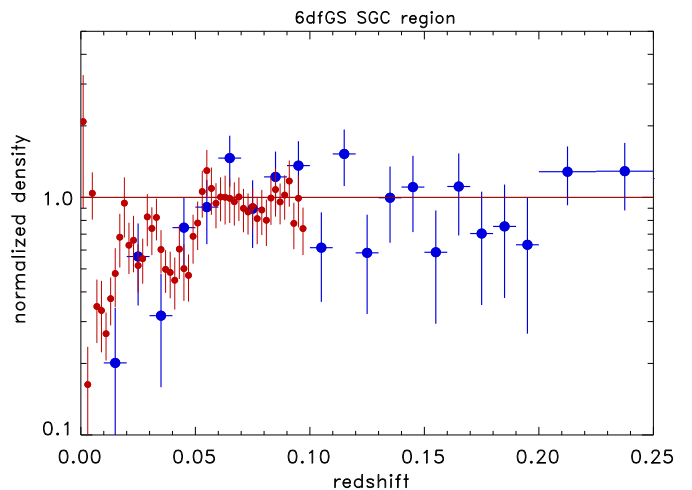}
\caption{{\sf REFLEX II} cluster density distribution as a function of redshift
for the region RA $= 0 - 50, 330 - 360^o$ 
and DEC $= -40 - 0^o$ in the south galactic cap
for a minimum luminosity of $0.2 \times 10^{44}$ erg s$^{-1}$ (0.1 - 2.4 keV).
The density distribution has been
corrected by means of the survey selection function and normalised to
unity. The region corresponds to region 6dFGS-SGC of Whitbourn \& Shanks (2014).
We also show the results of their assessment of the galaxy distribution 
out to $z = 0.1$ in this region with small red data points.
}\label{fig8}
\end{figure}

\begin{figure}
   \includegraphics[width=\columnwidth]{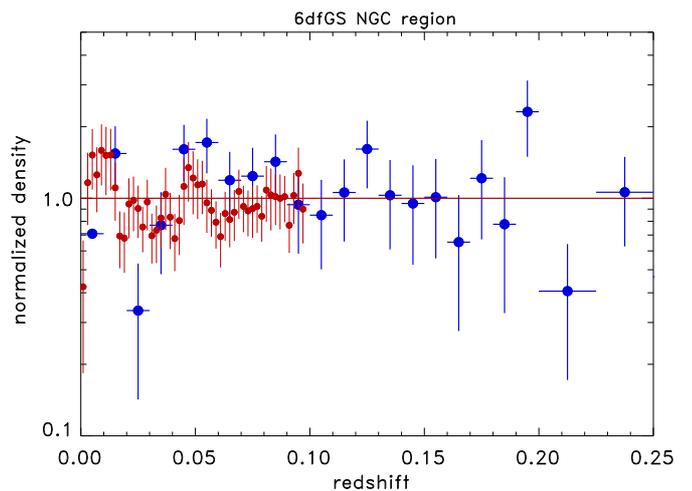}
\caption{{\sf REFLEX II} cluster density distribution as a 
function of redshift for the region RA $= 150 - 220^o$ and 
DEC $= -50 - 0^o$  in the north galactic cap for the same luminosity
limit as in Fig. 8. The density distribution has been
corrected by means of the survey selection function and normalised to
unity. The region corresponds to region 6dFGS-NGC of Whitbourn \& Shanks (2014).
We also show the results of their assessment of the galaxy distribution 
out to $z = 0.1$ in this region with small red data points.
}\label{fig9}
\end{figure}

%_________________________________One column table----------------------------
   \begin{table}
      \caption{Comparison of the ratio of the observed to 
predicted cumulative number of clusters as a function of the redshift limit 
for the {\sf REFLEX II} clusters in the 6dFGS-SGC and 6dFGS-NGC regions.
We also give the statistical uncertainty in brackets.}
         \label{Tempx}
      \[
         \begin{array}{lll}
            \hline
            \noalign{\smallskip}
  {\rm z~limit} & {\rm SGC}  & {\rm NGC} \\
            \noalign{\smallskip}
            \hline
            \noalign{\smallskip}
0.02  &  0.17 (\pm 0.12) & 1.41 (\pm 0.41) \\
0.03  &  0.37 (\pm 0.12) & 0.86 (\pm 0.22) \\
0.04  &  0.35 (\pm 0.10) & 0.83 (\pm 0.18) \\
0.05  &  0.45 (\pm 0.10) & 1.02 (\pm 0.17) \\
0.06  &  0.54 (\pm 0.09) & 1.15 (\pm 0.16) \\
0.07  &  0.69 (\pm 0.10) & 1.16 (\pm 0.15) \\
0.08  &  0.72 (\pm 0.09) & 1.17 (\pm 0.14) \\
0.10  &  0.84 (\pm 0.09) & 1.18 (\pm 0.12) \\
            \noalign{\smallskip}
            \hline
            \noalign{\smallskip}
         \end{array}
      \]
%\begin{list}{}{}
%\item[$^{\rm a}$] ......
%\end{list}
%{\bf Notes:} 
\label{tab1}
   \end{table}
%
%________________________Table-END______________________________________

\section{Discussion}

As discussed in the literature described above, an underdense region 
may become cosmologically interesting in terms of mimicking 
the observed accelerated expansion if the underdense 
region has a minimum size of $\sim 300$ Mpc and if our location is 
within 10\% of the center of the void (e.g. Alexander et al. 2009).
A first inspection of the redshift distribution of the {\sf REFLEX} clusters
indicates that the cluster density is on average underdense in the {\sf REFLEX}
region at $z \le 0.04$ which corresponds to a comoving radius 
of only $\sim 170$ Mpc. Thus the detected underdensity is significantly
smaller than what is required for the above void models.
Several tests showed that this underdensity is not an artifact of the
X-ray luminosity function used for the analysis. The homogeneity of the
density distribution on larger scales, as observed in the redshift range 
up to $0.3$ confirms that the observed density variations are 
confined to smaller volumes. This $z \le 0.3$  survey region 
corresponds to a maximum distance of 1.19 Gpc and a survey 
volume of 2.4 Gpc$^3$. In comparison, the size of the
volume in which we observe the cluster distribution to be significantly
underdense is 0.007 Gpc$^3$. Thus the {\sf REFLEX} survey shows
an almost homogeneous Universe on scales larger than about 300 Mpc.
This latter result supports the picture that the {\sf REFLEX II} survey
is large enough to constrain the extent of a local void.

In the next step of our analysis we probed the cluster distribution 
in more detail, and find that the underdensity in the local southern 
sky is mostly confined to the regions of the South Galactic Pole 
and the SGC at low galactic latitudes. The good correspondence of 
the galaxy and cluster distribution in the latter area and in the 
NGC region gives further support to the reliability of our measurements. 
These results also show, however, that we are not located in an 
isotropic underdense region. Therefore we conclude that the local 
large-scale structure traced by {\sf REFLEX II} does
not support a local large-scale void conforming to the minimum void models.
Most of the works that claim a detection of a local void are dominated
by galaxy data preferentially from regions in the SGC near the
equator and near the South Galactic Pole (Huang et al. 1997, Frith
et al. 2003, 2006, Busswell et al. 2004, Keenan et al. 2013, Whitbourn
\& Shanks 2014). These regions show local underdensities in our 
studies while other sky regions are not underdense at low
redshifts.

For a more quantitative characterisation of the local underdensity
we have to consider that galaxy clusters have a biased distribution
with respect to the matter distribution. The biasing factor is a function
of cluster mass and in a sample of clusters it depends on the 
biasing averaged over the mass distribution. For the {\sf REFLEX} cluster sample
the average bias has been modelled as a function of the lower
luminosity limit (Balaguera-Antolinez et al. 2011, Chon et al. 2014).
For the redshift range $z = 0 - 0.04$ the lower luminosity limit
varies from $L_x(min) = 10^{42}$ erg s$^{-1}$ to 
$L_x(min) = 6 \times 10^{42}$ erg s$^{-1}$ and the biasing factor is in
the range 2.5 to 3. The theoretical concept of cluster biasing has
been verified with {\sf REFLEX} with uncertainties smaller than the statistical
uncertainties in the present study.
Thus for an underdensity in the cluster distribution
of about $40 (\pm 15)\%$ we expect the matter distribution to have 
a smaller underdensity of about $15 (\pm 5)\%$.

Also for a quantitative comparison to the galaxy distribution we 
have to take this bias into account, since galaxies are hardly biased.
Table 3 lists the cumulative cluster density inside various redshifts
normalised to the expectations for the 6dF-SGC and 6dF NGC region.
We note that for $z \le 0.04$ the cluster density is lower 
in the 6dF-SGC area by about
$60(\pm 22)\%$ including systematic errors. This corresponds to an 
unbiased underdensity of about $22(\pm 8)\%$. The underdensity in the
galaxy distribution as noted in Fig. 8 is about $40 \%$, a bit
higher than our prediction, but still in reasonable agreement given
the large statistical uncertainties.

\section{Summary and Conclusion}

With the well understood selection function of the  {\sf REFLEX II}
galaxy cluster sample, we can trace the large-scale density 
distribution over a large region in an unbiased way.
Here we focussed on the study of the local density distribution
in the southern sky.

We traced the cluster distribution in the extragalactic 
sky to a redshift of $z \sim 0.3$, corresponding to a comoving
distance of $\sim 1.2$ Gpc and a survey volume of 2.4 Gpc$^3$,
and find the cluster distribution remarkably homogeneous
on large scales. With this result we can establish a large
scale mean cluster density, which allows us a precise
assessment of more local density variations and to confine their
extent well within our survey volume.

For the redshift range $z = 0$ to $0.04$ we find the cluster 
density to be lower than the large-scale average by about 30 - 40\%
with a significance larger than 3.4 $\sigma$. With the expected 
bias of the cluster density variations compared to those of
the dark matter, we infer an average dark matter underdensity
of about $15(\pm 5)\%$ in the southern extragalactic sky. 
This underdensity translates into a locally larger Hubble constant 
of about $\sim 3(\pm 1)\%$ assuming  $\Lambda$CDM cosmology.

A closer inspection
shows that this is not an isotropic underdensity in the southern sky.
A special region that contributes most to the observed underdensity
is located in the South Galactic Cap within {\sf REFLEX II}.

A comparison of the density distribution of galaxies and
clusters in the South and North Galactic Cap in the southern 
sky with data coming from Whitbourn \& Shanks (2014)
and from our {\sf REFLEX}
survey shows that both object populations trace the same density
distribution. In both surveys that SGC shows an underdensity
at redhifts from $z = 0$ to $0.05$ and no local underdensity in
the NGC.

Observations of a local underdensity in the galaxy distribution
on scales up to 300 Mpc have given rise to intense speculations
that the cosmic acceleration deduced from the observations of SN 
type Ia can be explained
by a local void region. Our results show that the local underdensity
is significantly smaller than required by minimum void models
and a locally underdense region is not observed in all directions
in the southern sky. 

\begin{acknowledgements}
H.B. and G.C. acknowledge support from the DFG Transregio Program TR33
and the Munich Excellence Cluster ''Structure and Evolution of the Universe''.  
G.C. acknowledges support by the DLR under grant no. 50 OR 1305.
M.B. acknowledges support of an STFC studentship and C.A.C  acknowledges
support through ARI's Consolidated Grant ST/J001465/1.
\end{acknowledgements}

\end{document}